\documentclass[12pt,draft]{article}

\usepackage{amssymb}
\usepackage{latexsym}

\newtheorem{theorem}{Theorem}[section]

\newtheorem{lemma}{Lemma}[section]

\newcommand{\nc}{\newcommand}
\nc{\C}{{\mathbb C}}
\nc{\R}{{\mathbb R}}
\nc{\HH}{{\mathbb H}}
\nc{\Z}{{\mathbb Z}}
\nc{\N}{{\mathbb N}}
\nc{\dd}{{\rm d}}
\nc{\DD}{{\rm D}}

\begin{document}

\title{The topology of asymptotically locally flat gravitational instantons}

\author{G\'abor Etesi
\\ {\it Department of Geometry, Mathematical Institute, Faculty of
Science,}
\\ {\it Budapest University of Technology and Economics,}
\\{\it Egry J. u. 1 H \'ep.,}
\\{\it H-1111 Budapest, Hungary}
\\ {\tt etesi@math.bme.hu}}

\maketitle

\pagestyle{myheadings}
\markright{G. Etesi: The topology of ALF gravitational instantons}

\thispagestyle{empty}

\begin{abstract}
In this letter we demonstrate that the intersection form of the 
Hausel--Hunsicker--Mazzeo compactification of a four dimensional 
ALF gravitational instanton is definite and diagonalizable over the integers
if one of the K\"ahler forms of the hyper-K\"ahler gravitational 
instanton metric is exact. This leads to their topological classification.

The proof exploits the relationship between $L^2$ cohomology and U$(1)$ 
anti-instantons over gravitational instantons recognized by Hitchin. We 
then interprete these as reducible points in a singular 
SU$(2)$ anti-instanton moduli space over the compactification providing
the identification of its intersection form.

This observation on the intersection form might be a useful tool in the 
full geometric classification of various asymptotically locally 
flat gravitational instantons.
\end{abstract}
\vspace{0.1in}
\centerline{Keywords: {\it Gravitational instantons, Yang--Mills 
instantons}}
\centerline{PACS numbers: 02.40.Ma; 04.90.+e; 11.15.-q}

\section{Introduction}

By a {\it gravitational instanton} we mean a connected, four dimensional 
complete hyper-K\"ahler Riemannian manifold. In particular these spaces 
have SU$(2)\cong {\rm Sp}(1)$ holonomy consequently 
are Ricci flat hence solve the Euclidean Einstein's vacuum equation. 
Since the only compact four dimensional hyper-K\"ahler spaces up to universal 
covering are diffeomorphic to the flat torus $T^4$ or a $K3$ surface, 
for further solutions we have to seek non-compact examples. 
Compactness in this case is naturally replaced by the condition that 
the metric be complete and decay to the flat metric somehow at infinity.

Such open examples can be constructed as follows. Consider a 
connected, orientable compact four-manifold $\overline{M}$ with connected 
boundary $\partial\overline{M}$ which is a smooth three-manifold. Then 
the open manifold $M:=\overline{M}\setminus\partial\overline{M}$ has a 
decomposition $M=C\cup N$ where $C$ is a compact subset and 
$N\cong\partial\overline{M}\times\R^+$ is 
an open annulus or neck. Parameterize the half-line $\R^+$ by $r$. 
Assume $\partial\overline{M}$ admits a smooth fibration 
$\partial\overline{M}\rightarrow B$ over a base manifold $B$ with fibers 
$F$. We suppose the complete hyper-K\"ahler metric $g$ can be written 
asymptotically and locally as
\begin{equation}
\dd r^2+r^2g_B +g_F.
\label{aszimptotika}
\end{equation}
In other words the base $B$ of the fibration blows up locally in a 
Euclidean way as $r\rightarrow\infty$ while the volume of the fiber remains 
finite. By the curvature decay $g_F$ must be flat hence $F$ is 
a connected, compact, orientable flat manifold. Depending 
on the dimension of $F$ we can introduce several cases of increasing 
transcendentality, using the terminology of Cherkis and Kapustin \cite{che}:
\begin{itemize}
\item[(i)] $(M,g)$ is ALE (asymptotically locally Euclidean) 
if $\dim F=0$;

\item[(ii)] $(M,g)$ is ALF (asymptotically locally flat) if $\dim F=1$,
in this case necessarily $F\cong S^1$ must hold;

\item[(iii)] $(M,g)$ is ALG (this abbreviation by induction) if $\dim 
F=2$, in this case $F\cong T^2$;

\item[(iv)] $(M,g)$ is ALH if $\dim F=3$, in this case $F$ is 
diffeomorphic to one of the six flat orientable three-manifolds.

\end{itemize}
Due to their relevance in quantum gravity or recently rather in 
low-energy supersymmetric solutions of string theory and last but 
not least their mathematical beauty, there has been some effort to classify 
these spaces over the past decades. Trivial examples for all cases are 
provided by the spaces $\R^{4-\dim F}\times F$ with their product flat 
metrics. The first two non-trivial infinite families in a rather explicit 
form were discovered by Gibbons and Hawking in 1976 \cite{gib-haw}. One of 
these families are the $A_k$ ALE or multi-Eguchi--Hanson spaces. In 1989 
Kronheimer gave a full classification of ALE spaces 
\cite{kro} constructing them as minimal resolutions of $\C^2/\Gamma $ 
where $\Gamma\subset{\rm SU}(2)$ is a finite subgroup i.e., $\Gamma$ is 
either a cyclic group $A_k$, $k\geq 0$, dihedral group $D_k$ with $k>0$, or 
one of the exceptional groups $E_l$ with $l=6,7,8$.

The other infinite family of Gibbons and Hawking is the $A_k$ ALF or 
multi-Taub--NUT family. Recently another $D_k$ ALF family has been 
constructed by Cherkis and Kapustin \cite{che-kap2} and 
in a more explicit form by Cherkis and Hitchin \cite{che-hit}. 

Using string theory motivations, recently Cherkis and 
Kapustin have been suggesting a classification scheme for ALF cases as 
well as for ALG and ALH \cite{che}. They claim that the $A_k$ and 
$D_k$ families with $k\geq 0$ exhaust the ALF geometry (in this 
enumeration $D_0$ is the Atiyah--Hitchin manifold) while for ALG it is 
conjectured that the possibilities are $D_k$ with $0\leq k\leq 5$ 
\cite{che-kap1} and $E_l$ with $l=6,7,8$. Conjecturally, there is only 
one non-trivial ALH space. The trouble is that these spaces are more 
transcendental as $\dim F$ increases hence their 
constructions, involving twistor theory, Nahm transform, etc. are less 
straightforward and explicit.

To conclude this brief survey we remark that by the restrictive 
hyper-K\"ahler assumption on the metric which appeared to be relevant 
in the later string theoretic investigations, some examples considered 
as ``gravitational instantons'' in the early eighties, have been 
excluded later. Non-compact examples which satisfy the above 
fall-off conditions are for instance the Euclidean Schwarzschild 
solution or the Euclidean Kerr--Newman solution which 
are complete ALF, Ricci flat but not hyper-K\"ahler spaces \cite{haw}. 
For a more complete list of such ``old'' examples cf. \cite{egu-gil-han}.
  
From Donaldson theory we have learned that the moduli spaces of 
SU$(2)$ instantons over compact four-manifolds encompass lot of 
information about the original manifold hence understanding SU$(2)$ 
instantons over gravitational instantons also might be helpful in their 
classification. The full construction of SU$(2)$ instantons
in the ALE case was carried out by Kronheimer and Nakajima in 1990 
\cite{kro-nak}. However already in the ALF case we run into 
analytical difficulties and have only sporadic 
examples of explicit solutions (cf. e.g. \cite{ali-sac}\cite{ete-hau1}
\cite{ete-hau2}) not to mention the ALG and ALH geometries.

An intermediate stage between gravitational instantons and their SU$(2)$ 
instanton moduli is $L^2$ cohomology because $L^2$ harmonic 2-forms always 
can be regarded as U$(1)$ (anti-)instantons hence reducible points in the 
(anti-)instanton moduli spaces over gravitational instantons. 
In spite of SU$(2)$ instantons, $L^2$ harmonic forms are quite well 
understood mainly due to the recent paper of Hausel--Hunsicker--Mazzeo 
\cite{hau-hun-maz}. Their construction reduces the calculation of the 
$L^2$ cohomology groups of $(M,g)$, highly non-trivial analytical 
objects, to the intersection hence topological cohomology of a certain 
compactification $X$ of $(M,g)$. A consequence of this is
the stability of the $L^2$ cohomology under compact perturbations of the
metric $g$. We will investigate this space $X$ 
in more detail below. For more historical remarks on $L^2$ cohomology we 
refer to the bibliography of \cite{hau-hun-maz}.

In this letter we focus our attention to the special case of ALF 
spaces. For this class $X$ is a connected, compact, orientable, smooth 
four-manifold without boundary. Put an orientation onto $X$ by 
extending the canonical orientation of $M$ coming from the hyper-K\"ahler 
structure. Assume in this moment for simplicity that the metric also extends 
conformally to $X$ that is we can construct an oriented Riemannian manifold 
$(X,\tilde{g})$ whose restriction to $M$ is conformally equivalent to $(M,g)$. 
Assume furthermore that there exists a K\"ahler form on $(M,g)$ 
satisfying $\omega =\dd\beta$. Then the ALF condition implies that 
$\beta$ has linear growth and $L^2$ harmonic 2-forms are anti-self-dual by a 
theorem of Hitchin \cite{hit} hence, if suitable normed, represent 
U$(1)$ anti-instantons over $(M,g)$. Moreover they extend as reducible
SU$(2)$ anti-instantons over $(X,\tilde{g})$ by a codimension 2
singularity removal theorem of Sibner and Sibner \cite{sib-sib}, reporved 
by R\aa de \cite{rad}. Assume finally that $X$ is simply connected. Then 
$L^2$ harmonic 2-forms provide all the theoretically possible
reducible points in the moduli space. We will demonstrate here using elements 
of Hodge theory and Donaldson theory that this implies definiteness and 
diagonalizability of the intersection form of $X$ over the integers. 

This observation leads to the topological classification of ALF gravitational 
instantons subject to the above technical conditions using Freedman's 
fundamental theorem \cite{fre}. It turns out 
that if $k$ denotes the dimension of the 2nd $L^2$ cohomology space of 
$(M,g)$ then $X$ is homeomorphic either to the four-sphere if $k=0$ or 
to the connected sum of $k>0$ copies of complex projective spaces with 
reversed orientation.     

\section{$L^2$ cohomology and anti-self-duality}

Remember that a $k$-form $\varphi$ over an oriented Riemannian manifold 
$(M,g)$ is called an {\it $L^2$ harmonic $k$-form} if 
$\Vert\varphi\Vert_{L^2(M,g)}<\infty$ 
and $\dd\varphi =0$ as well as $\delta\varphi =0$ where $\delta =-*\dd 
*$ is the formal adjoint of $\dd$ (and $*$ is the Hodge star operation). 
The space ${\cal H}^k(M,g)$ of all $L^2$ harmonic 
$k$-forms is denoted by $\overline{H}^k_{L^2}(M,g)$ and is called the 
{\it $k$th (reduced) $L^2$ cohomology group of $(M,g)$}. This is a natural 
generalization of the de Rham cohomology over compact manifolds to the 
non-compact case since by de Rham's theorem ${\cal H}^k(M,g)$ 
is isomorphic to the corresponding de Rham cohomology group $H^k(M)$ 
over a compact manifold.

Consider a non-compact gravitational instanton $(M,g)$ of any type 
(i)-(iv) above. If one tries to understand the $L^2$ cohomology 
of it, it is useful to consider a compactification $X$ of $M$ by 
collapsing the fibers $F_r$ over all points of $B_r$ 
as $r\rightarrow\infty$ \cite{hau-hun-maz} (remember that 
$\partial\overline{M}_r:=\partial\overline{M}\times\{ r\}$ is fibered over 
$B_r$ with fibers $F_r$, $r\in\R^+$). In general $X$ is not a 
manifold but a stratified space only with one singular stratum 
$B_\infty$ and a principal stratum $M=X\setminus B_\infty$. However if 
$F_r$ is a sphere, the resulting space $X$ will be a connected, compact, 
oriented, smooth four-manifold without boundary. If $(M,g)$ is an ALF 
gravitational instanton then $F_r\cong S^1$ therefore $X$ is smooth and 
$M$ is identified with the 
complement of the two dimensional submanifold $B_\infty\subset X$. The 
following theorem is a consequence of Corollary 1 in \cite{hau-hun-maz} 
combined with vanishing theorems of Dodziuk \cite{dod} and Yau 
\cite{yau} (also cf. Corollary 9 therein):

\begin{theorem}{\rm (Hausel--Hunsicker--Mazzeo, 2004)}
Let $(M,g)$ be a gravitational instanton of ALF type. 
Then we have the following natural isomorphisms
\[\overline{H}^k_{L^2}(M,g)\cong\left\{\begin{array}{ll}
                 H^k(X)  & \mbox{if $k=2$}, \\ [2mm]
                 0 & \mbox{if $k\not= 2$}
                                 \end{array}\right.\]
where the cohomology group on the right hand side is the ordinary 
singular cohomology. $\Diamond$
\label{hau-hun-maz}
\end{theorem}
We see then that for an ALF gravitational instanton 
$L^2$ cohomology reduces to degree 2. Hence we focus our 
attention to this case and re-interpret it as follows 
(cf. \cite{ete-hau1}). Over an oriented Riemannian four-manifold if 
$\varphi$ is an $L^2$ harmonic 2-form then so is its Hodge-dual
$*\varphi$; consequently $F^\pm :={{\bf i}\over 2}(\varphi\pm *\varphi)$ 
are (anti-)self-dual imaginary $L^2$ harmonic 2-forms. 
Over a contractible ball $U\subset M$ the condition $\dd F^\pm\vert_U =0$ 
implies the existence of a local 1-form $A^\pm_U$ such that
$F^\pm\vert_U =\dd A^\pm_U$. Clearly, if $U\cap V\not=\emptyset$ for two 
balls $U,V$ and $\dd A^\pm_U$, $\dd A^\pm_V$ are two representatives of 
$F^\pm$ on $U\cap V$ then $A^\pm_V=A^\pm_U+\dd\chi^\pm_{UV}$ holds for an 
imaginary function $\chi^\pm_{UV}$. If $F^\pm$ represents a non-trivial 
element of $H^2(M,\R )$ then we assume 
that $\left[{1\over 2\pi{\bf i}}F^\pm\right]\in H^2(M,\Z )$ by 
linearity yielding
$F^\pm$ always can be regarded as the curvature of an (anti)-self-dual 
U$(1)$ connection $\nabla^\pm$ of finite action with curvature 
$F_{\nabla^\pm}=F^\pm$ on a complex line bundle $L^\pm$ over $M$. 

Proceeding further $\nabla^\pm$ induces a connection $(\nabla^\pm )^{-1}$ 
with curvature $-F_{\nabla^\pm}$ on the dual bundle $(L^\pm )^{-1}$ 
providing a reducible SU$(2)$ connection $\nabla^\pm\oplus (\nabla^\pm )^{-1}$ 
on the splitted bundle $E^\pm=L^\pm\oplus (L^\pm )^{-1}$ with curvature 
$F^\pm\oplus (-F^\pm )$. We fix the normalization constants such that
\begin{equation}
-{1\over 8\pi^2}\int\limits_M {\rm tr}(F_{\nabla^\pm}\oplus 
(-F_{\nabla^\pm})\wedge F_{\nabla^\pm}\oplus (-F_{\nabla^\pm} 
))=-{1\over 4\pi^2}\int\limits_M
{\rm tr}(F_{\nabla^\pm}\wedge F_{\nabla^\pm}) =\pm 1.
\label{kvantalas}
\end{equation}   
Summing up, an $L^2$ harmonic 2-form $\varphi$ induces a reducible 
SU$(2)$ (anti-)instanton $\nabla^\pm\oplus (\nabla^\pm )^{-1}$ on a 
splitted SU$(2)$ bundle $E^\pm =L^\pm\oplus (L^\pm )^{-1}$ over $(M,g)$ 
with second Chern number $\pm 1$.

In case of an ALF gravitational instanton however, a more intrinsic 
relationship exists between $L^2$ cohomology and anti-self-duality which 
removes the disturbing self-dual-anti-self-dual ambiguity from this 
construction. To see this we will assume that there is an orthogonal complex 
structure $J$ on $(M,g)$ whose induced K\"ahler form, given by
\[\omega (X,Y)=g(JX, Y)\]
for two vector fields, satisfies $\omega =\dd\beta$ for a 1-form 
$\beta$. Consider the infinite neck $N\subset M$ and pick up a local 
unitary orthonormal frame $\{X_1, X_2, X_3, X_4\}$ 
on $U\subset N$ with respect to $J$ and $g$ that is, 
$g(X_i, X_j)=\delta_{ij}$ and $X_2=JX_1$ and $X_4=JX_3$. Take a 
diffeomorphism $\phi :N\cong\partial\overline{M}\times\R^+$ and denote 
by $r$ the coordinate on $\R^+$ as usual. Since
(\ref{aszimptotika}) describes a metric of asymptotically vanishing
curvature, then using the induced frame on $\phi (U)\subset 
\partial\overline{M}\times\R^+$ satisfying $\phi_*X_1=\partial 
/\partial r$ the metric and the almost complex structure locally decay as
\begin{equation}
(\phi^{-1})^*g =\pmatrix{1 & 0 & 0 & 0\cr 
                         0 & 1 & 0 & 0\cr
                         0 & 0 & 1 & 0\cr
                         0 & 0 & 0 & 1} +O(1/r^p),\:\:\:\:\:
(\phi^{-1})^*J =\pmatrix{ 0 & 1 & 0 & 0\cr 
                         -1 & 0 & 0 & 0\cr 
                           0 & 0 & 0 & 1\cr 
                          0 & 0 & -1 & 0} + O(1/r^q)
\label{lecsenges}
\end{equation}
for some positive exponents $p$ and $q$.
The induced K\"ahler form satisfies $\omega_{ij}=J_i^kg_{kj}$ and 
by integrating (\ref{lecsenges}) the 
corresponding $\beta$ has {\it linear growth} that is,
\[\vert (\phi^{-1})^*\beta\vert_g\leq c_0+c_1r\]
for positive constants $c_0$ and $c_1$ as $r\rightarrow\infty$. Since 
the metric is hyper-K\"ahler we can apply a theorem of Hitchin (cf. 
Theorem 4 in \cite{hit}):

\begin{theorem}{\rm (Hitchin, 2000)} Let $(M,g)$ be a four dimensional 
complete hyper-K\"ahler manifold with induced orientation from the 
complex structures. Assume one of the K\"ahler forms satisfies $\omega 
=\dd\beta$ where $\beta$ has linear growth. Then any $L^2$ harmonic form is 
anti-self-dual. $\Diamond$
\end{theorem}
Consequently over an ALF gravitational instanton satisfying the above 
technical assumption $\varphi$ can be identified 
with the curvature of an U$(1)$
anti-instanton $\nabla$ by writing $F_\nabla ={\bf i}\varphi$. We regard 
this as a reducible SU$(2)$ 
anti-instanton $\nabla\oplus\nabla^{-1}$ on the splitted bundle $E=L\oplus 
L^{-1}$ over $(M,g)$ with second Chern number $-1$. 

To proceed further, let us consider the extendibility of various fields 
from $M$ over $X$. The identification of $L^2$ harmonic 2-forms
with Yang--Mills connections can be exploited at this point. First we 
endow $X$ with an orientation induced by the orientation of $M$ used in 
Hitchin's theorem. Next consider the metric. Take a smooth positive 
function $f:M\rightarrow\R^+$ of suitable asymptotics, that is 
$f(r)\sim O(r^{-2})$, then the rescaled metric $\tilde{g}:=f^2g$ extends over 
$X$ as a symmetric tensor field. In other words, there is a symmetric tensor 
field $\tilde{g}$ on $X$ which is conformally equivalent to $g$ on 
$X\setminus B_\infty =M$ and degenerates along $B_\infty$ providing an 
oriented, degenerate Riemannian manifold $(X,\tilde{g})$. 

The Yang--Mills connection $\nabla\oplus\nabla^{-1}$ is also singular 
along $B_\infty\subset X$ which is a smooth, codimension 2 
submanifold. Consider an open neighbourhood $V_\varepsilon$ of 
$B_\infty$ in $X$. We suppose $V_\varepsilon$ is a disk bundle over 
$B_\infty$ and locally we can introduce polar coordinates $(\rho, \theta)$ 
with $0\leq\rho <\varepsilon$ and $0\leq\theta <2\pi$ on the disks over 
all points of $B_\infty$. Let $f_\varepsilon$ be a smooth positive 
function on $X$ which vanishes on $X\setminus V_\varepsilon$ and is equal to 
1 on $V_{{\varepsilon\over 2}}$. Finally choose a smooth Riemannian metric 
$h$ on $X$ and consider the regularization
\[\tilde{g}_\varepsilon :=f_\varepsilon h+(1-f_\varepsilon 
)\tilde{g}.\]
It is clear that $\tilde{g}=\tilde{g}_0$. The action is conformally 
invariant hence the original assumption $\Vert F_\nabla\oplus (-F_\nabla 
)\Vert_{L^2(M,g)}<\infty$ together with the completeness of $g$ implies 
that $F_\nabla\in L^2(X, \tilde{g}_\varepsilon )$ 
hence the perturbed norm $\Vert F_\nabla\oplus (-F_\nabla 
)\Vert_{L^2(V_\varepsilon , \tilde{g}_\varepsilon)}$ is arbitrary small 
if $\varepsilon$ tends to zero. Taking into 
account (\ref{kvantalas}), which rules out fractional holonomy of the 
connection along the disks, and referring to a codimension 2 
singularity removal theorem of Sibner and Sibner \cite{sib-sib} and R\aa 
de \cite{rad} one can show that the SU$(2)$ anti-instanton 
$\nabla\oplus\nabla^{-1}$ extends over $B_\infty$ as a smooth SU$(2)$ 
connection which is of course not anti-self-dual with respect to the 
regularized metric $\tilde{g}_\varepsilon$. Notice that this connection is 
independent of the metric $\tilde{g}_\varepsilon$. Then taking the limit 
$\varepsilon\rightarrow 0$ we recover a smooth reducible connection on a 
smooth SU$(2)$ vector bundle $E$ over $X$ satisfying $c_2(E)=-1$ which 
is {\it formally} anti-self-dual over $(X, \tilde{g}_0)$. 

Consider the space ${\cal B}_E(X):={\cal A}_E(X)/{\cal G}_E(X)$ of 
gauge equivalence classes of smooth SU$(2)$ connections on the fixed bundle 
$E$. Let ${\cal M}_E(X,\tilde{g}_\varepsilon )\subset{\cal B}_E(X)$ be 
the moduli of smooth SU$(2)$ anti-instantons on $E$ over 
$(X,\tilde{g}_\varepsilon )$. For a generic smooth $\tilde{g}_\varepsilon$ 
this is a smooth compact manifold except finitely many points. Then as 
$\varepsilon\rightarrow 0$ the
moduli space ${\cal M}_E(X,\tilde{g}_0)$ intersects in ${\cal B}_E(X)$ the 
reducible connections constructed above. The singularity of $\tilde{g}_0$ 
somewhat 
decreases the overall regularity of the moduli space however this fact will 
not cause any inconvenience in our considerations ahead. Summing up again it 
is clear by now that the cohomology class 
$[\varphi ]\in\overline{H}^2_{L^2}(M,g)$ is 
identified with a reducible point $[\nabla\oplus\nabla^{-1}]\in 
{\cal M}_E(X,\tilde{g}_0)$.

\section{Topological classification}

Now we are in a position to make two simple observations on the 
intersection form of $X$ hence the topology of $M$. We assume that $X$ is 
simply connected. Recall moreover that $X$ has a fixed orientation.

\begin{lemma} We find $b^+(X)=0$ that is, the intersection form $q_X: 
H^2(X,\Z )\times H^2 (X, \Z )\rightarrow\Z$ of $X$ is negative definite.
\end{lemma}
{\it Proof.} It follows on the one hand that 
\[\left[ {1\over 2\pi{\bf i}}F_{\nabla}\right]=c_1(L)\in H^2(X,\Z ).\]
On the other hand $F_{\nabla}$ is anti-self-dual hence harmonic 
yielding at the same time
\[{1\over 2\pi{\bf i}}F_{\nabla}\in {\cal H}^-(X, \tilde{g}_0)\]
where ${\cal H}^-(X, \tilde{g}_0)$ contains anti-self-dual harmonic
2-forms. However the decomposition 
\[{\cal H}^2(X, \tilde{g}_\varepsilon )={\cal H}^+(X, 
\tilde{g}_\varepsilon )\oplus {\cal H}^-(X, \tilde{g}_\varepsilon )\] 
into self-dual and anti-self-dual harmonic 2-forms
induces $H^2(X,\R )=H^+(X,\R )\oplus H^-(X,\R )$ through the 
Hodge isomorphism for $\varepsilon >0$ and $\dim H^\pm (X,\R )=b^\pm (X)$.
Consequently we get a similar decomposition of $H^2(X,\R )$ for 
$\varepsilon\rightarrow 0$.

It follows that the 
subspace ${\cal H}^-(X, \tilde{g}_0)\cong H^-(X,\R )\subset H^2(X,\R )$ 
depends on the metric $\tilde{g}_0$, has codimension $b^+(X)$ and 
contains as many as $b^2(X)=\dim H^2(X,\R )$ linearly 
independent points of $H^2(X, 
\Z )\subset H^2(X,\R )$ taking into account Theorem \ref{hau-hun-maz}.
Notice that $H^2(X,\Z )$ is a lattice of rank $b^2(X)$ because $X$ is 
simply connected. However ${\cal H}^-(X, \tilde{g}_0)\cap H^2(X, \Z )$ cannot 
contain $b^2(X)$ linearly independent points if $b^+(X)>0$. Hence 
$b^+(X)=0$ must hold that is, $q_X$ is negative definite. $\Diamond$
\vspace{0.1in}

\noindent Secondly we claim that
\begin{lemma} The intersection form $q_X$ of $X$ diagonalizes over the 
integers.
\end{lemma}
{\it Proof.} Fix an SU$(2)$ vector bundle $E$ over $X$ satisfying $c_2(E)=-1$. 
The gauge equivalence classes of reducible anti-self-dual 
connections on $E$ are in one-to-one correspondence with the possible 
decompositions $E=L\oplus L^{-1}$ i.e., writing $c_1(L)=\alpha$, with 
the set of topological reductions of $E$ over $X$:
\[{\cal R}_E(X)=\{ (\alpha, -\alpha )\:\vert\: \alpha\in H^2(X,\Z 
),\:\alpha^2=-1\}.\]
Consider the set $\{ (\alpha_1 ,-\alpha_1),\dots 
,(\alpha_k,-\alpha_k)\}$ of all possible topological reductions of $E$. 
First of all, we can see that $\alpha_1,\dots,\alpha_k$ are linearly 
independent in $H^2(X, \Z )$. Indeed, taking into account that the 
intersection form is negative definite, from the triangle inequality we 
get $-1<q_X(\alpha_i, \alpha_j)<1$ consequently $\alpha_i$ is 
perpendicular to $\alpha_j$ if $i\not= j$ hence linear independence 
follows.

Furthermore, notice that we have identified the reducible points 
with $L^2$ harmonic 2-forms therefore Theorem \ref{hau-hun-maz} shows 
$k=\dim H^2(X,\Z )$ and since $q_X(\alpha_i, \alpha_j)=-\delta_{ij}$ in 
this basis $q_X$ is diagonal. $\Diamond$
\vspace{0.1in}

\noindent This leads to our main theorem:

\begin{theorem} Let $(M,g)$ be an ALF gravitational 
instanton satisfying $\omega =\dd\beta$ for one of its K\"ahler form. 
Let $X$ be the Hausel--Hunsicker--Mazzeo compactification of $(M,g)$. 
Assume $X$ is simply connected and is endowed with an orientation induced 
by any of the complex structures on $(M,g)$. Let 
$k=\dim\overline{H}^2_{L^2}(M,g)$. Then there is a 
homeomorphism of oriented topological manifolds
\[X\cong S^4\]
if $k=0$ or
\[X\cong\underbrace{\overline{\C P}^2\#\overline{\C 
P}^2\#\dots\#\overline{\C P}^2}_{k}\]
if $k>0$ where $\overline{\C P}^2$ is the complex projective space with 
reversed orientation (compared with the orientation induced by its standard 
complex structure).
\end{theorem} 
{\it Proof.} $X$ is simply connected and its 
intersection form is negative definite by the previous lemma 
consequently $q_X\cong 0$ if $k=0$. Hence $q_X$ is
isomorphic to the intersection form of $S^4$. The theorem for 
$k=0$ then follows from Freedman's classification of
simply connected topological four-manifolds \cite{fre}. Similarly, 
$q_X\cong <-1>\oplus\dots\oplus <-1>$ ($k$ times) if $k>0$. Hence $q_X$ is 
isomorphic to the intersection form of the connected sum of $k$ copies of 
$\overline{\C P}^2$'s hence $q_X$ is of odd type. But note that $X$ has 
a smooth structure as well as the connected sum of $k$ copies of 
$\overline{\C P}^2$'s. Applying again Freedman's result we obtain the 
theorem for the remaining cases. $\Diamond$
\vspace{0.1in}

\noindent{\it Remark.} The case $X\cong S^4$ is realized by the flat space 
$\R^3\times S^1$ while the compactification of the multi-Taub--NUT space 
with as many as $k$ NUTs provides an example for the remaining case.

The hyper-K\"ahlerity of the metric is important in our
construction. For example, in case of the Euclidean Schwarzschild
manifold there are two independent $L^2$ harmonic
2-forms, one is self-dual while the other is anti-self-dual
\cite{ete-hau1}. Anti-self-dualizing them one cancels out hence cannot
be detected in the anti-instanton moduli space. Notice that this geometry 
is not hyper-K\"ahler and the intersection form of its compactification, 
which is $S^2\times S^2$, is indefinite.

\section{Concluding remarks}
We conjecture that all ALF gravitational instantons satisfy the 
technical condition $\omega =\dd\beta$ (in fact it is true for all 
known examples) however we cannot prove this in this moment.
\vspace{0.1in}

\noindent{\bf Acknowledgement.} The work was supported by OTKA grants 
No. T43242 and No. T046365.


\begin{thebibliography}{99}

\bibitem{ali-sac} Aliev, A.N., Saclioglu, C.: {\it Self-dual fields 
harbored by a Kerr--Taub-BOLT instanton}, Phys. Lett. {\bf 
B632}, 725-727 (2006);

\bibitem{che} Cherkis, S.A.: {\it Self-dual gravitational 
instantons}, Talk given at the AIM-ARCC workshop ``$L^2$ 
cohomology in geometry and physics'', Palo Alto, USA, March 16-21 (2004);

\bibitem{che-hit} Cherkis, S.A., Hitchin, N.J.: {\it Gravitational 
instantons of type $D_k$}, Commun. Math. Phys. {\bf 260}, 299-317 
(2005);

\bibitem{che-kap1} Cherkis, S.A., Kapustin, A.: {\it Hyper-K\"ahler 
metrics from periodic monopoles}, Phys. Rev. {\bf D65}, 084015 (2002);

\bibitem{che-kap2} Cherkis, S.A., Kapustin, A.: {\it $D_k$ 
gravitational instantons and Nahm equations}, Adv. Theor. Math. Phys. {\bf 2}, 
1287-1306 (1999);

\bibitem{dod} Dodziuk, J.: {\it Vanishing theorems for square-integrable 
harmonic forms}, Proc. Indian Acad. Sci. Math. Sci. {\bf 90}, 21-27 
(1981);

\bibitem{egu-gil-han} Eguchi, T., Gilkey, P.B., Hanson, A.J.: {\it 
Gravity, gauge theories and differential geometry}, Phys. Rep. {\bf 66}, 
213-393 (1980);
 
\bibitem{ete-hau1} Etesi, G., Hausel, T.: {\it Geometric interpretation 
of Schwarzschild instantons}, Journ. Geom. Phys. {\bf 37}, 126-136 
(2001);

\bibitem{ete-hau2} Etesi, G., Hausel, T.: {\it On Yang--Mills instantons 
over multi-centered gravitational instantons}, Commun. Math. Phys. {\bf 
235}, 275-288 (2003);

\bibitem{fre} Freedman, M.H.: {\it The topology of four-manifolds}, Journ. 
Diff. Geom. {\bf 17}, 357-454 (1982);

\bibitem{gib-haw} Gibbons, G.W., Hawking, S.W.: {\it Gravitational 
multi-instantons}, Phys. Lett. {\bf B78}, 430-432 (1976);

\bibitem{hau-hun-maz} Hausel, T., Hunsicker, E., Mazzeo, R.: {\it Hodge 
cohomology of gravitational instantons}, Duke Math. Journ. {\bf 122}, 
485-548 (2004);

\bibitem{haw} Hawking, S.W.: {\it Gravitational instantons}, Phys. Lett.
{\bf A60}, 81-83 (1977);

\bibitem{hit} Hitchin, N.J.: {\it $L^2$ cohomology of hyper-K\"ahler 
quotients}, Commun. Math. Phys. {\bf 211}, 153-165 (2000);

\bibitem{kro} Kronheimer, P.B.: {\it The construction of ALE spaces as 
hyper-K\"ahler quotients}, Journ. Diff. Geom. {\bf 29}, 665-683;

\bibitem{kro-nak} Kronheimer, P.B., Nakajima, N.: {\it Yang--Mills 
instantons on ALE gravitational instantons}, Math. Ann. {\bf 288}, 
263-307 (1990);

\bibitem{rad} R\aa de, J.: {\it Singular Yang--Mills fields. Local theory 
II.}, Journ. Reine Angew. Math. {\bf 456}, 197-219 (1994);
 
\bibitem{sib-sib} Sibner, I.M., Sibner, R.J.: {\it Classification of 
singular Sobolev connections by their holonomy}, Commun. Math. Phys. 
{\bf 144}, 337-350 (1992);

\bibitem{yau} Yau, S-T.: {\it Harmonic functions on complete Riemannian 
manifolds}, Commun. Pure. Appl. Math. {\bf 28}, 201-228 (1975).
 
\end{thebibliography}
\end{document}